\documentstyle[12pt,aaspp4]{article}
\begin{document}
\title{ Large Scale QSO-galaxy correlations for radio loud 
and optically selected QSO samples} 
\author{N. Ben\' \i tez\altaffilmark{1} \& E. Mart\' \i nez-Gonz\'alez}
\affil{Instituto de F\' \i sica de Cantabria, CSIC-Universidad de Cantabria,\\
Facultad de Ciencias, Avda. Los Castros s/n, 39005 Santander, Spain} 
\altaffiltext{1}{Dpto. de F\' \i sica Moderna, Universidad de Cantabria,
Facultad de Ciencias, Avda. Los Castros s/n, 39005 Santander, Spain}  
\begin{abstract}
We have studied the distribution of $B_J<20.5$ galaxies from the ROE/NRL 
COSMOS/UKST catalogue around two samples of $z>0.3$ QSOs with
similar redshift distributions. The first sample is formed by 144 
radio-loud QSOs from the Parkes Catalogue, and the other contains 167 
optically selected QSOs extracted from the Large Bright Quasar Survey. 
It is found that there is a $\approx 
99.0\%$ significance level excess of COSMOS/UKST galaxies around the 
PKS QSOs, whereas there is a marginal defect of galaxies around the 
LBQS QSOs. When the distribution of galaxies around both samples is 
compared, we found that there is an overdensity of galaxies around the 
PKS sample with respect to the LBQS sample anticorrelated with the 
distance from the QSOs at a $99.7\%$ significance level. Although this 
result apparently agrees with the predictions of the multiple 
magnification bias, it is difficult to explain by gravitational 
lensing effects alone; dust in the foreground galaxies and selection 
effects in the detection of LBQS QSOs should be taken into account. 
It has been established that the lines of sight 
to PKS flat-spectrum QSOs go through significatively 
higher foreground galaxy densities than the directions to LBQS quasars, 
what may be partially related with the reported reddening of PKS QSOs.
\end{abstract}
\section{Introduction}\label{intr}

In the last few years, several studies have established the existence
of a statistical excess of line-of-sight companion galaxies around 
high redshift quasars. Although it has been suggested
that these objects belong to clusters or groups which are physically 
associated to the quasars (\cite{hin91}; \cite{tys86}), in order to be 
detected at such high redshifts they should be undergoing strong 
luminosity evolution. This seems unlikely on the light of the 
recent data on galaxy evolution obtained through the study of 
absorption-selected galaxy samples (\cite{ste95}), which shows that 
the most plausible (and often the unique) interpretation 
for many of these observations is the existence of a magnification 
bias caused by gravitational lensing (see the reviews 
\cite{sch92}; \cite{sch95}; \cite{wu95}).

The density of a population of flux-limited background sources 
(e.g. QSOs) behind a gravitational lens is affected by the lens 
magnification $\mu$ in two opposite ways. 
One of the effects is purely geometrical: as the angular 
dimensions of a lensed patch of the sky are expanded by a 
factor $\mu$, the physical size of a region observed through a 
fixed angular aperture will be smaller than in the absence of the lens. 
Because of this, the QSO surface density will 
decrease by a factor $\mu$ with respect to the unlensed background density
\cite{Na89}).
On the other hand, the lens will magnify faint quasars (which would not 
have been detected otherwise ) into the sample and increase the number
of detected QSOs (\cite{can81}; \cite{sch86}, etc.). 
If the slope of the quasar number-counts cumulative distribution is 
steep enough, this effect would dominate over the angular 
area expansion and there would be a net excess of QSOs behind the lens. 
Foreground galaxies trace the matter overdensities acting as lenses and 
thus there will be a correlation between the position in the sky of these 
galaxies (or other tracers of dark matter as clusters) and the background 
quasars. This QSO-galaxy correlation is characterized by the overdensity 
factor $q$ (\cite{sch89}), which is defined as the ratio of the 
QSO density behind a lens with magnification $\mu$ to the unperturbed 
density on the sky. Its dependence on the effective slope $\alpha$ of 
the QSO number counts distribution (which has the form 
$n(>S)\propto S^{-\alpha}$, or $n(<m)\propto 10^{0.4\alpha m}$) 
and the magnification $\mu$ can be expressed as (\cite{Na89})
\begin{equation}
q \propto \mu^{\alpha-1}
\end{equation}
We see that the value of q critically depends on the number counts slope 
of the background sources. For instance, if the number counts are shallow 
enough, ($\alpha < 1$), there would be negative QSO-galaxy associations. 
It is clear that in order to detect strong, positive QSO-galaxy 
correlations due to the magnification bias, we have to use QSO 
samples with very steep number counts slopes.
\cite{bo91} have shown that for a QSO sample which 
is flux-limited in two bands (with uncorrelated fluxes), 
$\alpha$ is substituted by $\alpha_{eff}$, the sum of the number 
counts-flux slopes in those bands. This effect is called 
'double magnification bias'. Since $\alpha_{eff}$ is 
usually $>1$ for samples which are flux-limited in both the 
optical and radio bands (e.g. radio-loud QSOs), a strong
positive QSO-galaxy correlation should be expected for them.

It is important to understand when QSO samples may be affected by
the double magnification bias. The usual identification procedure for 
a X-ray or radio selected QSO sample involves setting up a flux 
threshold in the corresponding band and obtaining follow-up optical 
images and spectra of the QSO candidates. The observer is limited 
in this step by several circumstances (e.g. the sensitivity of the detectors 
or the telescope diameter), and even if the QSO sample was not intended to
be optically selected, in the practice there will be an
optical flux threshold for the QSOs to enter the sample. 
Therefore the existence of an explicit and homogeneus flux-limit 
in the optical band is not as essential for the presence of the 
magnification bias as the vaue of the effective slope of the unperturbed 
number counts. 
If this slope is steep enough, the effect should be detectable even 
in incomplete samples, and often more strongly than in complete 
catalogues: within such samples, the optically 
brightest QSOs (i.e., those more likely to be lensed) are usually 
the first to be identified, as they are easier to study 
spectroscopically or through direct imaging. 

At small angular scales, ($\theta\lesssim$ few $\arcsec$) the existence of 
QSO-galaxy correlations is well documented for several QSO samples obtained 
with different selection criteria (\cite{web88}; see also \cite{ha90} and 
\cite{wu95} for reviews). As expected due to the double magnification
bias effect, the correlations are stronger for radio-loud quasars 
(\cite{tho94}). In the cases where no correlation is 
found, e.g. for optically-selected and relatively faint quasars, 
the results are still consistent with the magnification bias effect and 
seem to be due to the shallowness of the QSO number counts distribution 
at its faint end (\cite{wu94}). 

\cite{ha90} reviewed the studies on QSO-galaxy correlations 
(on both small and large scales). After assuming that the galaxies 
forming the excess are physical companions to the QSO, they showed that 
while the amplitude of the radio-quiet QSO-galaxy correlation quickly
declines at $z\gtrsim 0.6$, the inferred radio-loud QSO-galaxy
correlation function steadily increases with redshift, independently
of the limiting magnitude of the study. It should be noted that
such an effect will be expected, if a considerable part of the galaxy 
excess around radio-loud QSOs is at lower redshifts. 
If a foreground group is erroneously considered to be physically 
associated with a QSO, the higher the redshift of the QSO, the
stronger the 3-D clustering amplitude that will be inferred. This source 
of contamination should be taken into account carefully when carrying 
out these studies, as there is evidence that part of the detected excess 
around high redshift radio-loud QSOs is foreground and related with the 
magnification bias.

The association of quasars with foreground galaxies on scales 
of several arcmin may arise as a consequence of lensing by the 
large scale structure as proposed by \cite{ba91} (see also 
\cite{sch95} and references therein). Several authors have also 
investigated QSO-cluster correlations: \cite{roho94}, \cite{ses94} 
and \cite{wuha95}. There are not many studies of large scale QSO-galaxy 
correlation, mainly because of the difficulty in 
obtaining apropriate, not biased, galaxy samples. Besides, the 
results of these studies may seem contradictory, as they radically 
differ depending on the QSO type and limiting magnitude. 

\cite{boy88} found a slight anticorrelation between the positions of 
optically selected QSOs and COSMOS galaxies. When they cross-correlated
the QSOs with the galaxies belonging to clusters, the anticorrelation 
become highly significant ($4\sigma$) on $4\arcmin$ scales. Although 
this defect was interpreted as due to the presence of dust in 
the foreground clusters which obscured the quasars, the recent 
results of \cite{mao95} and \cite{lrw95} have imposed limits on the 
amounts of dust in clusters which 
seem to contradict this explanation. It seems more natural, 
taking into account the rather faint flux-limit of the QSOs of 
\cite{boy88} to explain this underdensity as a result of the 
magnification bias effect.
 
\cite{smi95} do not find any excess of foreground, $B_J< 20.5$ APM 
galaxies around a sample of $z>0.3$ X-ray selected QSOs. It should be
noted that although the objects in this sample are detected in two bands 
(X-ray and optical), the expected excess should not be increased by 
the double magnification bias effect, because the X-ray 
and optical fluxes are strongly correlated (\cite{boy96}). For these QSOs,
the overdensity should be roughly the same as for the optically selected 
ones.

On the other hand, there have been strong evidences of positive large
scale associations of foreground galaxies with high redshift
radio-loud QSOs. As e.g., between the radio-loud quasars from the 1Jy 
catalogue (\cite{sti94}) and the galaxies from the Lick (\cite{fug90}, 
\cite{bar93}), IRAS (\cite{bbs96}, \cite{bar94}) and APM (\cite{ben95a}) 
catalogues. In the latter paper, we found that APM galaxies are 
correlated with the 1Jy QSO positions, but did not have enough statistics 
to reach the significance levels reported in the present paper. 
Unlike the COSMOS 
catalogue, the APM catalogue provides magnitudes in two filters, $O$(blue) 
and $E$(red). When we considered only the reddest, $O-E>2$ galaxies, 
the overdensity reached a significance level of $99.1\%$.
Afterwards, \cite{ode95} showed (using a similar, but
much more processed catalog, the APS scans of POSS-I) that the 
galaxies which trace the high-density cluster and filamentary regions
have redder $O-E$ colors than the galaxies drawn from low-density
interfilamentary regions. If the fields containing Abell clusters were
removed from the sample of \cite{ben95a}, the results did not change
significantly, so it seems that the detected effect was 
caused by the large scale structure as a whole.
\cite{fort} confirmed the existence of gravitational lensing 
by the large scale structure by directly detecting the large 
foreground invisible matter condensations responsible for the 
magnification of the QSOs through the polarization produced by 
weak lensing in several 1Jy fields. 

These differences in the nature of the QSO-galaxy associations 
for the several QSO types seem to arise quite naturally when we 
take into account the effects of the double magnification bias. 
There is not any strong correlation between the radio and optical 
fluxes for radio-loud quasars, so for these objects the overdensity 
will be $\propto \mu^{\alpha_{opt}+\alpha_{rad}-1}$ (\cite{bo91}), 
where $\alpha_{opt}$ and $\alpha_{rad}$ are the number-counts slope 
in the optical and in radio respectively. If we assume that 
$\alpha_{opt}$ is roughly the same for radio-loud and optically 
selected QSOs (although this is far from being clear), the overdensity 
of foreground galaxies should be higher for the radio-loud QSOs. For 
optically and X-ray selected samples (because of the X-ray-optical 
flux correlation ), $\alpha_{eff}$ and therefore the overdensity, will 
be smaller. 

In any case, it is difficult to compare conclusively the published
analyses of QSO-galaxy correlations.
They have been performed using galaxy samples obtained with
different filters and magnitude limits, and which therefore 
may not trace equally the matter distribution because of 
color and/or morphological segregation or varying depths. 
Besides, the QSO samples differ widely in their redshift 
distributions. As the magnification of a background source depends on 
its redshift, QSO samples with different redshift distributions will 
not be magnified equally by the same population of lenses.
It would thus be desirable, in order to disentangle these factors 
from the effects due to the magnification bias caused by gravitational
lensing, to reproduce the mentioned studies using galaxy samples 
obtained with the same selection criteria and QSO samples which 
are similarly distributed in redshift. This is the purpose of the 
present paper: we shall study and compare the distribution of 
COSMOS/UKST galaxies around two QSO samples, one radio-loud and the 
other radio-quiet with practically identical redshift distributions. 

It is also interesting to mention in this context the results of 
\cite{web95}. These authors observed that a sample of flat-spectrum 
radio-loud quasars from the Parkes catalogue (\cite{pks}) displays a 
wide range of $B_J-K$ colours, apparently arising from the presence 
of varying amounts of dust along the line of sight. Optically selected 
quasars do not present such a scatter in $B_J - K$ and lie at the lower 
envelope of the $B_J - K$ colours, so Webster and collaborators 
suggested that
the selection criteria used in optical surveys miss the reddest
quasars. Although there seems to be several other, more plausible,
reasons for this phenomenon ( see for instance \cite{boy96}), it is
not adventurous to suppose that it may be partially related to the
differences between the foreground projected environments of
radio-loud and optically selected quasars. If a considerable part of
the absorbing dust belongs to foreground galaxies, a greater density
of these galaxies would imply more reddening.

The structure of the paper is the following: Section 2 describes the 
selection procedures of both QSO samples and galaxy fields
and discuss the possible bias which could have been introduced in the data.
Section 3 is devoted to the discussion of the statistical methods
used in the literature for the study of this problem and applies them
to our data. Section 4 discusses the results obtained in Sec 3. 
Section 5 lists the main conclusions of our work.

\section{The Data}\label{data}

As we have explained above, the aim of our paper is the study of the
distribution of foreground galaxies 
around two QSO samples, one radio-loud and the other radio quiet, 
which have similar redshift distributions. It would also be interesting
to find out if these differences in the foreground galaxy density 
occur for the QSO samples of \cite{web95}. This, as 
we have mentioned, could be related to the differential reddening of
the radio-loud quasars. Therefore, in order to form a radio-loud sample 
suitable for our purposes but as similar as possible to the one 
used by \cite{web95}, we collect all the quasars from the PKS catalogue 
which are listed in the \cite{ver96} QSO catalogue and obey the 
following constraints: a) flux $>$ 0.5Jy at 11cm; b) $-45 < \delta 
< 10$ and 
c) galactic latitude $|b| > 20$. So far, we do not constrain the radio 
spectral index of the quasars. This yields a preliminary sample of 276 
quasars.

 The galaxy sample is taken from the ROE/NRL COSMOS/UKST Southern Sky
object catalogue, (see \cite{yen92} and references therein). It contains 
the objects detected in the COSMOS scans of the glass copies of the 
ESO/SERC blue survey plates. The UKST survey is organized in $6\times6$ 
square degree fields on a 5 degree grid and covers the zone 
$-90 < \delta <0$ excluding a $\pm 10$ deg zone in the galactic plane. 
COSMOS scans cover only $5.4\times 5.4$ square degrees of a UKST field.
The scan pixel size is about 1 arcsec. Several parameters are supplied 
for each detected object, including the centroid in both sky and plate 
coordinates, $B_J$ magnitude and the star-galaxy image 
classification down to a limiting magnitude of $B_J\approx21$. 

We are going to study the galaxy distribution in $15 \arcmin $ radius 
fields centered on the quasars of our sample. Due to several factors
as vignetting and optical distorsions, the quality of the object
classification and photometry in Schmidt plate based surveys
degrades with increasing plate radius. Therefore, 
we constrain our fields to be at a distance from the plate center of 
$r=\sqrt{\Delta x^2 +\Delta y^2} < 2.5$ degrees. Besides, to avoid 
the presence of fields which extend over two plates we further
restrict our sample to have $|\Delta x_k|, |\Delta y_k| < 2.25$ degrees, 
where $\Delta x_k$ and $\Delta y_k$ are the distances, in the $\alpha$ 
and $\delta$ directions respectively, from the center of the plate 
( because of the UKST survey declination limits, this also 
constrains our quasar sample to have $\delta < 2.25$). 
After visually inspecting all the fields, several of them (6) 
are excluded from the final sample 
because they present meteorite traces. We end up with 147 circular
fields with a $15\arcmin$ radius
centered on an equal number of Parkes Quasars. This subsample 
of radio-loud quasars is, as far as we know, not biased towards 
the presence of an excess of foreground galaxies, which is the essential
point for our investigation. 
In order to avoid contamination from galaxies physically 
associated with the quasars, we also exclude three $z<0.3$ 
quasars from our radio-loud sample (\cite{smi95} point out that only 
$5\%$ of $B_J < 20.5$ galaxies have $z>0.3$), 
which is finally formed by 144 fields.

We have excluded a region of $12\arcsec$ around the 
positions of the quasars (which may have an uncertainty up to 
$\pm 5$ arcsec). 
This is done to avoid the possibility of counting the quasar as a galaxy 
(there are six cases in which the quasar is classified as an 
extended object) and because of the great number of 'blended' 
objects at the quasar position. Most of these pairs of objects are 
classified as 'stars' when deblended, but taking into account 
the pixel resolution, it would be desirable to examine the original 
plates or, better yet, perform high resolution CCD imaging in order 
to properly deblend and classify these objects as many of them could 
be galaxies.

The optically selected sample is taken from the Large Bright Quasar 
Survey as in \cite{web95}. This prism-selected catalogue contains 1055 
quasars brighter than $B_J \approx 18.6$ on several equatorial and 
southern fields (for details see \cite{hew90}).
In order to form an optically selected subsample of quasars
we have begun by choosing the 288 quasars from the LBQS catalogue
which were closest in redshift to our final sample of 144 PKS quasars.
We impose to them exactly the same constraints in sky and plate 
position as to the radio-loud quasar fields. Finally we 
visually examined the fields and excluded six of them because of 
meteorite 
traces. The resulting number of fields is 167. As the LBQS extends over
relatively small areas of the sky, several of these fields overlap. 
We have
checked that their exclusion from the statistical tests performed below
does not affect significantly the result, so we leave them in the 
sample.

The resulting redshift distribution for both QSO samples is plotted 
in Fig 1. A Kolmogorov-Smirnov test cannot distinguish between them
at a 94.5\% significance level. We merge all the fields 
in each sample into two 'superfields' which contain all the objects 
classified as galaxies with $B_J<20.5$. This is a reasonable limiting 
magnitude, and has been already used by other investigators 
(\cite{smi95}). The PKS merged field 
contains 15235 galaxies whereas the LBQS field only contains 14266. 
This is a difference of $24\%$ in the average object background density, 
well over a possible Poissonian fluctuation, and seems to be caused 
by the presence of misclassified stars in our galaxy sample at low 
galactic latitudes. The Parkes fields extend over
a much wider range of galactic latitudes $(|b| > 20^o)$ than the LBQS
ones, which are limited to high galactic latitudes $(|b|>45^o)$ and
thus much less affected. In fact, we measure 
the existence of a significant anticorrelation 
between the absolute value of the galactic latitude $|b_k|$ of the fields 
and the total number of objects in each field $N_{gal}$. The correlation 
is still stronger between $N_{gal}$ and sec$(90-|b|$), as shown in Fig. 2,
with a correlation coefficient $\rho=0.4, (p>99.99\% )$.  
This contamination should be randomly distributed over the field
and would lower the significance of any possible correlation and 
make it more difficult to detect. In order to check this, we have 
correlated the overdensity $n_{in}/n_{out}$ of objects within the inner 
half of every
individual field, ($n_{in}$ is the number of objects within the
inner half of the field surface and $n_{out}$ is the number of
objects in the outer half) with sec$(90^o-|b|)$,  as can be seen
in Fig. 3.  If anything, there might be a slight anticorrelation 
(the Spearman's rank correlation test only
gives a significance of $80\%$) in the sense that the fields
with more star contamination are the ones which show less excess of 
galaxies in the inner half of the field. This is what could 
be expected if there were a
genuine galaxy excess close to the QSO positions; this excess should
be diluted by the presence of stars randomly distributed with respect
to the QSOs. Excluding the fields with $|b|\leq 30^o$, as in
\cite{smi95}, does not change significantly the main results, 
as we show in Fig 4.
Because of this contamination by stars, there is a slight bias in 
the data which makes harder to detect QSO-galaxy correlations for the PKS 
QSOs than for the LBQS ones. We have also checked that there are no other 
correlations between $N_k$ and $q_k$ and other possible relevant 
parameters as the plate or sky position of the fields. 

\section {Statistical Analysis}\label{stats}

The study of QSO-galaxy correlations due to the 
magnification bias effect is complicated by several circumstances.
The amplitude of the correlation function $w_{qg}$ is expected to
be rather weak, and strongly dependent on the limiting magnitude 
of the galaxies and the QSOs. Besides, the shape of $w_{qg}$ as 
a function of $\theta$ is unknown ( it seems that the interesting theoretical 
estimation of \cite{ba94} has not been confirmed empirically 
by \cite{bbs96}). 

In the past, several methods have been used to detect and statistically 
establish the existence of these correlations. 
One of the most simple and widespread approaches has consisted in counting 
the number of galaxies $N_{in}$ in a somehow arbitrarily 
defined region centered on the quasars and comparing the value found 
with its statistical expectation, which is measured either from the outer 
regions of the fields or from some other comparison fields which are 
assumed to have the same density of galaxies on average. The significance 
can be inferred empirically (\cite{ben95a}) 
or just by considering that N has a poissonian 
distribution with $\sqrt{N}$ rms. This last assumption seems to hold well
in some cases, when the number of individual fields is very large, 
but for other samples, usually smaller, the rms is found to be 
$\alpha\sqrt{N}$, where $\alpha\approx 1.1-1.5$ (\cite{ben95b}). 
A shortcoming of this method is that it does not extract 
all the information contained in the fields as it only explores the 
distribution of galaxies around the quasar in a small number of scales, 
and often uses the rest of the field just to determine the average density.
Besides, if the correlation scale is comparable with the dimensions of
the fields, the average density measured on the fields would be increased 
with respect to the unperturbed one and thus an artificially lowered 
signification
will be obtained. 

Another method, the rank-correlation test was used in \cite{bar93},
\cite{bar94}. All the individual galaxy fields are merged into a 
single 'superfield', which is subsequently divided into
$N_{bins}$ annular intervals of equal surface. A Spearman's rank 
order test is applied to determine if the number of
galaxies in each bin $n_i,( i=1,N_{bins})$ is anticorrelated with the 
bin radius $r_i$. This test does not require any assumption about the 
underlying probability distribution of the galaxies and takes into account
all the information contained in the fields. However it has several drawbacks.
The rings have all equal surface, so we end up with more 
bins in the outer parts of the fields, which are less sensitive 
from the point of view of detecting the correlation. Besides, the method 
only 'senses' the relative ordering of $n_i$ and $r_i$ over the whole field. 
For instance if 
$w_{qg}(\theta)$ is very steep and goes quickly to zero, there will be only 
a few bins clearly over the average in the central region, and the 
correlation coefficient could then be dominated by the more numerous 
outer bins with nearly no excess galaxies. 
The value of the correlation coefficient and its significance, depend thus
critically on the number of chosen bins and the dimension of the 
fields. However, this test can still be useful if the correlation 
scale is similar to the scale of the field. 

Recently, Bartsch et al. (1996) have introduced the weighted-average test. 
They define the estimator $r_g$ as 
\begin{equation}
r_g={1\over N}\sum^N_{j=1}g(\theta_j),  
\end{equation}
where N is the total number of galaxies in the merged field, and $\theta_j$
are the radius from the QSO of each galaxy. They show, under certain
general assumptions, that if the galaxies
in the merged field are distributed following a known QSO-galaxy 
correlation function $w_{gq}(\theta)$, for $g(\theta) \propto w_{gq}(\theta)$ 
the quantity $r_g$ is optimized to distinguish such a distribution of 
galaxies from a random one. They take 
$w_{gq}(\theta)=(0.24+h\theta/deg)^{-2.4}$ from the theoretical results 
of \cite{ba94} ($h=H_o/100$ Mpc km s$^{-1}$), 
and show with simulated fields that this method 
is slightly more sensitive than Spearman's 
rank order test. However,
when they study the correlation between IRAS galaxies and the QSOs from
the 1Jy catalogue with the weighted-average test they obtain a much higher 
significance for their result than using the rank order test. 
They conclude that although the IRAS galaxies do not seem to be 
clustered around the QSOs following Bartelmann's correlation function, 
the weighted average method seems to be a much more efficient estimator than 
the rank order test. 
This is not surprising if we consider that, when calculating the 
estimator $r_g$ (as long as we use a steep enough form for $g(\theta)$) 
this test gives a much higher weight to the galaxies closer to the QSO,
that is, to the regions where the excess signal-to-noise is higher. An 
extreme case would be to use a top hat function with a given width 
$\theta_o$ as $g(\theta)$ (which would be equivalent to counting 
galaxies in a central circle of dimension $\theta_o$). This 
arbitrariness in the choice of $g(\theta)$ when we do not know 
the real shape of the QSO-galaxy correlation is a drawback 
of this method. Another problem is that the probability distribution of 
$r_g$ is unknown a priori. Because of that, the significance has to be 
determined using simulations, and as we have seen before, the real galaxy 
distribution is not always easy to know and model. 
Nevertheless, when we know theoretically the correlation, this test 
should be optimal, and it may also be useful in many other cases.

We have applied a variant of the rank order test to study the
distribution of galaxies around our PKS and LBQS samples. We also use 
the Spearman's rank order test as statistical estimator (in the 
implementation of \cite{nrec}), but instead of fixing a number of 
bins and dividing the field in rings of equal surface as in \cite{bar94}, 
the variables to be correlated will be $w(\theta_i)$ and $\theta_i$, 
where $w(\theta_i)$ is the value of the empirically determined 
angular correlation function in rings of equal width and 
$\theta_i$ is the distance of the bins from the QSO. Now, in general, 
each ring will have a different number of galaxies, but the values 
of $\theta_i$ are going to be uniformly distributed in radius, 
and thus we will not give more weight to the outer regions of the field.
As a statistical estimator we shall use $Z_d$, the number of 
times by which $D$, the so-called sum squared difference of ranks, deviates
from its null-hypothesis expected value. $D$ has an approximately 
normal distribution and is defined as
\begin{equation}
D=\sum^N_{i=1}(R_i-S_i)^2
\end{equation}
where $R_i$ is the rank of the radius of the i-th ring and $S_i$ is
the rank of the density in that same ring. Trying to avoid the 
dependence of the result on the concrete number of bins, we have 
followed this procedure: we have chosen a minimal ring width 
($0.4\arcmin$) in order to have at least $\approx 20$ galaxies in the first 
ring, and a maximal width ($0.75\arcmin$) 
so that there are at least 20 rings within the field. Then we perform 
8 different binnings changing the ring width in steps of $0.05\arcmin$, 
estimate $Z_d$ for each of them and calculate its average $<Z_d>$. 
This estimator should be very robust as it does not depend so strongly 
on the concrete value obtained for a binning, and the significance can be 
estimated directly from the value of $Z_d$ without the need of simulations.
The value for the PKS field is $<Z_d>=2.33\sigma$, $p=99.0\%$ and for the 
LBQS field $<Z_d>=-0.68\sigma$, $p=75.2\%$. 
We have also confirmed this by estimating $<Z_d>$ for $10^5$ simulations
with randomly distributed galaxies for each of both fields: the empirical 
significance for the PKS field is $p=99.01\%$ whereas the LBQS field gives 
$p=72.46\%$. This similarity in the values of the probabilities also 
confirms that the distribution of the galaxies in the fields is practically 
Poissonian. The QSO-galaxy correlation function for the PKS and LBQS 
sample is shown in Fig. 4 and 5 respectively. Error bars are poissonian
and the bin width is $0.75\arcmin$. In Fig. 4 we also show, without
error bars, the correlation function obtained for the PKS fields with
$|b|>30^o$

 In order to further test our results, we have also applied Bartsch et al. 
(1996) test to our data using Bartelmann's $g(\theta)$ and have 
estimated the significance with $10^5$ simulated fields for each 
sample with the same number of galaxies as the real fields randomly 
distributed. 
Strictly speaking this is an approximation, as the galaxies are clustered 
among themselves, but we have studied the number of galaxies on rings 
of equal surface (excluding a few central rings) and their distribution 
is marginally consistent with being Gaussian with a rms $\approx \sqrt{N}$, 
what is not surprising if we take into account the great number of fields 
contributing to each bin. 
The existence of a positive QSO-galaxy correlation for the PKS sample 
is detected at a significance level of $98.85 \%$. 
On the other hand, when we apply this test to the LBQS 
merged field, a slight anti-correlation is found at a level of 
$88.85\%$. These results are comparable to the ones obtained with the 
previous test. We have also tried other arbitrarily chosen variants of 
the function $g(\theta)$ to see the dependence of significance of 
the PKS excess on the concrete shape of $g(\theta)$: a Gaussian 
with a $2\arcmin$ width (analogous to a box of this size) yields $p=99.66\%$ 
and a $\propto \theta^{-0.8}$ law ( the slope of the 
galaxy-galaxy correlation function) gives $p=99.5\%$.
We see that for our galaxies, the shape of $g(\theta)$ proposed 
by Bartelmann is not optimal, and the significance depends 
sensitively on the shape of the function. However, tinkering with the form 
of $g(\theta)$ may be dangerous, as it could lead to creating an 
artificially high significance if we overadjust the shape of the function 
to the data. 

Thus, it seems that there is a positive QSO-galaxy correlation
in the PKS fields, and what appears to be a slight anticorrelation in the
LBQS ones. In order to measure how much these two radial distributions
differ, we have performed a series of binnings as the one 
described above for our test and defined $q_{i}$ in each ring as 
$q_i\propto n^{PKS}_i/n^{LBQS}_i$, where $n^{PKS}_i$ is the number 
of objects in each ring of the PKS field and $n^{LBQS}_i$ is the number of 
objects in the same bin of the LBQS field, and normalize by the mean 
density of each field. 
We apply the rank order test to all the resulting sequences of 
$q_i$ and bin radii $r_i$ as described
above and find that $<Z_d>=2.77$, $p=99.7\%$. $10^5$ simulations of field 
pairs give a significance of $p=99.74$. This means that the
density of foreground galaxies around the radio-loud quasars is higher
than in front of the optically selected sample, and is anticorrelated with
the distance from the QSOs at a $99.7\%$ significance level.

\section{Discussion}\label{disc}

As shown above, we have confirmed the existence 
of large scale positive correlations between high-redshift radio-loud 
QSOs and foreground galaxies, whereas for optically selected QSOs with 
the same redshift distribution the correlation is null or even negative. 
Can these results be explained by the double magnification bias
mentioned in the introduction? In order to answer this question 
the value of the number-counts distribution slopes in expression (1)
must be determined. These slopes can be estimated from the 
empirical distribution functions of our QSO samples.
The cumulative number-radio flux distribution for the PKS QSOs is plotted
in Fig. 6. A linear squares fit gives an approximate slope 
$\alpha^{PKS}_{rad} \approx 1.6$. The histogram of the distribution 
of $B_J$ magnitudes for the LBQS and the PKS QSOs is plotted in Fig 7a.
For the PKS QSOs we do not use the magnitudes quoted in 
\cite{ver96} as they have been obtained with different filters and
photometric systems. Instead, we have obtained $B_J$ magnitudes 
from the ROE/NRL COSMOS/UKST catalog, which should be reasonably 
homogeneous and accurate for $16<B_J<20.5$, apart from the intrinsic 
variability of the QSOs. Fig. 7a shows that PKS QSOs extend over 
a wider range of 
magnitudes than the LBQS ones, which have $B_J \lesssim 18.6$.
In Fig 7b we show 
the cumulative distributions of both QSO samples, $N(<B_J)$ as a 
function of $B_J$. The LBQS distribution (crosses) can be
well aproximated by a power law $\propto 10^{0.4\alpha^{LBQS}_{opt}}$ 
with $\alpha^{LBQS}_{opt}\approx 2.5$. 
The PKS distribution (filled squares) is more problematic
and cannot be approximated reasonably by a single power law. Although 
at brighter magnitudes it seems to have a slope similar to the LBQS ones,
it quickly flattens and has a long tail towards faint magnitudes.
Due to the incompleteness of the PKS sample, this distribution 
can be interpreted in two ways: either the flattening 
is caused by the growing incompleteness at fainter optical magnitudes and  
the slope of the underlying unperturbed distribution for the radio 
loud QSOs is the same as for LBQS ones, or the distribution function 
is intrisically shallow, and we are approximately observing its true form. 
Fortunately this is not a critical question; as it will be shown below, 
the difference between the slope values obtained in both cases is not 
enough to change significantly our main conclusions about the causes 
of the overdensity. Then, we roughly estimate the optical 
slope for the PKS distribution function with a linear 
squares fit between $16 < B_J < 17.75$ which yields  
$\alpha^{PKS}_{opt}\approx 1.9$. 
These slopes imply an overdensity of galaxies 
around the LBQS and PKS QSOs 
\begin{eqnarray}
&q^{LBQS} =\mu^{\alpha_{opt}^{LBQS}-1}\approx \mu^{1.5}\nonumber\\
&\\
&q^{PKS} =\mu^{\alpha_{opt}^{PKS}+\alpha_{rad}^{PKS}-1}\approx \mu^{2.5}\nonumber
\end{eqnarray}
That is, $q^{PKS}/q^{LBQS}\approx\mu$. At e.g. $\theta=2\arcmin$, 
for the LBQS we found $q^{LBQS}=0.968\pm0.063$. This yields a value for
the magnification $\mu = 0.98 \pm 0.04$. Then for the PKS QSOs, the 
overdensity should be $\approx 0.95 \pm 0.1$. However at $\theta=2\arcmin$, 
we measure $q_{PKS}=1.164\pm 0.061$. If we assume that the intrinsic 
PKS $B_J$ number-counts slope is the same as for the LBQS QSOs, 
$\alpha_{opt}^{PKS}=2.5$, we still cannot make both overdensities 
agree with a same magnification. 
In order to obtain these results with 'pure' gravitational lensing, a 
slope $\alpha_{PKS}^{opt}> 4$ would be needed. For 
smaller scales, the situation does not change, since $q^{PKS}/q^{LBQS}$ 
is still higher. Therefore, we must conclude that it is unlikely 
that the multiple magnification bias alone explains the results found. 

As mentioned above, some authors have explained the optically
selected QSO-cluster anticorrelations as due to the existence of dust 
in clusters (\cite{mao95} and references therein). What would be 
the expected overdensity when we consider the combined effects of 
magnification and dust absorption? Let's consider a patch of the sky 
which has an average magnification of $\mu$ for background sources and an 
average flux extinction of $\tau$ for a given optical band, i.e. 
the observed flux $S$ from the background sources in that band would be 
$S\approx(1-\tau)S_o$, where $S_o$ is the flux that we would measure in the 
absence of absorption. If we consider that the radio emission suffers no 
attenuation by the dust, the overdensity estimations for our samples would be 
\begin{eqnarray}
&q^{PKS}=\mu^{\alpha_{opt}^{PKS}+\alpha_{rad}^{PKS}-1}(1-\tau)^
{\alpha_{opt}^{PKS}}
\approx \mu^{2.5}(1-\tau)^{1.9}\nonumber\\
&\\
&q^{LBQS}=\mu^{\alpha_{opt}^{LBQS}-1}(1-\tau)^{\alpha_{opt}^{LBQS}}
\approx \mu^{1.5}(1-\tau)^{2.5}\nonumber
\end{eqnarray}
Therefore, from our results on scales of $2\arcmin$ we find 
$\mu\approx 1.139$, and $\tau\approx 0.089$. This extinction is 
consistent with the results of \cite{mao95}.
Although these values should be considered only as rough estimations, 
they show that considering dust absorption together with the multiple
magnification bias produces new qualitative effects in the
behavior of the overdensities of the different QSO types.
The strength of the overdensity is attenuated in both samples 
of QSOs, but the effect is stronger for the LBQS QSOs, 
which have a steeper optical counts slope. If we consider that 
the dust approximately traces the matter potential wells acting as lenses, 
i.e. that there is a strong correlation between magnification and 
extinction, the QSOs which are more magnified are also the ones which are
more absorbed. However, if the product $\mu(1-\tau)$ is greater than
unity, the net effect for each QSO will be a flux increment.

An alternative explanation is the existence of the bias 
suggested by \cite{rm92} and \cite{mao95}. They interpret that 
the avoidance of foreground clusters by optically selected QSOs is 
probably a selection effect due to the difficulty in identyfing quasars 
in crowded fields. In that case, apart from the slight 
QSO-galaxy anticorrelation generated by this effect, 
the LBQS samples would avoid the zones where the lensing by the large 
scale structure is stronger and thus their average magnification $\mu$ 
would be smaller than that of the PKS, which would not be affected by 
this selection bias. Besides, if dust and magnification are correlated,
radio-loud QSOs would also be more reddened on average than optically 
selected QSOs.

Regarding flat-spectrum QSOs, if we set an additional constraint for
our QSOs, $\gamma > -0.5$, where $\gamma$ is the slope of the 
spectral energy distribution, the resulting sample of 107 $z>0.3$ QSOs 
should be a fair representation of the radio-loud sample used by 
\cite{web95} for the study of reddening in QSOs. We apply again our rank 
correlation test to the field obtained by merging these 107 fields and 
conclude that the COSMOS/UKST galaxies are correlated with flat-spectrum
QSOs at a $98.5\%$ level. The QSO-galaxy correlation function is plotted 
in Fig. 8 with $0.75 \arcmin$ bins. The value of the overdensity 
is similar, but as we have now fewer fields, the significance level is lower. 
Nevertheless, if we take into account the small 
amounts of dust allowed by the observations of \cite{mao95}, it seems very
unlikely that all the reddening measured by \cite{web95} for the PKS QSOs
is due to dust absorption by foreground galaxies, although in some cases 
this effect could contribute considerably, as it has been shown by 
\cite{sti96}. This question could be clarified by cross-correlating 
the reddening of the QSOs with the density of foreground galaxies on 
small scales.

\section{Conclusions}

 We have studied the clustering of galaxies from the ROE/NRL 
COSMOS/UKST catalogue up to 15 $\arcmin$ scales around two QSO
samples with $z>0.3$. One of them contains 144 radio-loud QSOs from 
the Parkes Catalogue, and the other is formed by 167 optically selected 
QSOs obtained from the Large Bright Quasar Survey. 

There is a $\approx 99.0\%$ significance level excess of COSMOS 
$B_J<20.5$ galaxies around the PKS QSOs, whereas there is a slight 
defect of galaxies around the LBQS QSOs. We have compared the distribution 
of galaxies around both samples, and 
found that there is an overdensity around the PKS sample with respect 
to the LBQS sample anticorrelated with the distance from the QSO at a $99.7\%$ 
significance level. Whilst this result could be thought to agree 
qualitatively with the theoretical predictions of the multiple 
magnification bias effect, we show that it is difficult to 
explain it through gravitational lensing effects alone, and dust in the 
foreground galaxies and/or selection effects in the detection of LBQS 
QSOs must be considered. 

Finally, we have established that the lines of sight to 
PKS flat-spectrum QSOs go through significantly higher foreground 
galaxy densities than the directions to LBQS quasars. This may be 
related, at least partially, with the reddening of the PKS 
QSOs observed by \cite{web95}.
\acknowledgements 

The authors acknowledge financial support from the Spanish DGICYT, 
project PB92-0741. NB acknowledges a Spanish M.E.C. Ph.D. fellowship.
The authors are grateful to Tom Broadhurst, Jos\'e Luis Sanz 
and Ignacio Ferreras for carefully reading the manuscript and 
making valuable suggestions, and Sebastian Sanchez for useful
comments. They also thank D.J. Yentis for his help.

\newpage
\figcaption{Redshift distributions of the PKS (solid line histogram) 
and LBQS (dashed line histogram) QSOs in our samples. The vertical axis 
show the fraction of QSOs in each redshift bin. }
\figcaption{ The total number of objects classified as galaxies in each
field, $N_{gal}$, versus sec$(90^o-|b|)$, 
where $b$ is the field galactic latitude.
The continuous line represents the linear least squares fit, with a 
correlation coefficient $\rho=0.4$ and a significance $p>99.99\% )$.
Crosses and filled squares represent LBQS and PKS QSOs respectively.} 
\figcaption{$n_{in}/n_{out}$ vs. sec$(90^o-|b|)$. $n_{in}/n_{out}$ 
 is the ratio between the number of galaxies in the inner half of 
each field $n_{in}$ and the number of galaxies in the outer half $n_{out}$.
$b$ is the galactic latitude of each field.}
\figcaption{QSO-galaxy correlation for the PKS sample (continuous line). 
Error bars are poissonian and the bin width is approximately 
$0.75\arcmin$. The dotted line, with no error bars, represents the 
correlation function for a subsample of the PKS QSOs with $|b|>30^o$.}
\figcaption{QSO-galaxy correlation for the LBQS sample. 
Error bars are poissonian and the bin width is approximately $0.75\arcmin$.}
\figcaption{Logarithmic plot of the cumulative number counts--radio flux
distribution $N(>S)$ for the PKS QSOs (dotted line). The continuous 
line is a linear least squares fit which yields a slope 
$\alpha^{PKS}_{rad}=1.58$}
\figcaption{a) Histograms of the magnitude distribution for the PKS (solid
line) and LBQS (dashed line) QSOs. b) Cumulative number counts magnitude plot 
for the LBQS (crosses) and PKS (filled squares) QSOs as a function of 
$B_J$. The LBQS distribution is fitted by a power law 
$\propto 10^{0.4\alpha^{LBQS}_{opt}}$ with $\alpha^{LBQS}_{opt}\approx 2.5$.} 
\figcaption{QSO-galaxy correlation for the flat spectrum QSOs in the 
PKS sample. Error bars are poissonian and the bin width is $0.75\arcmin$.}
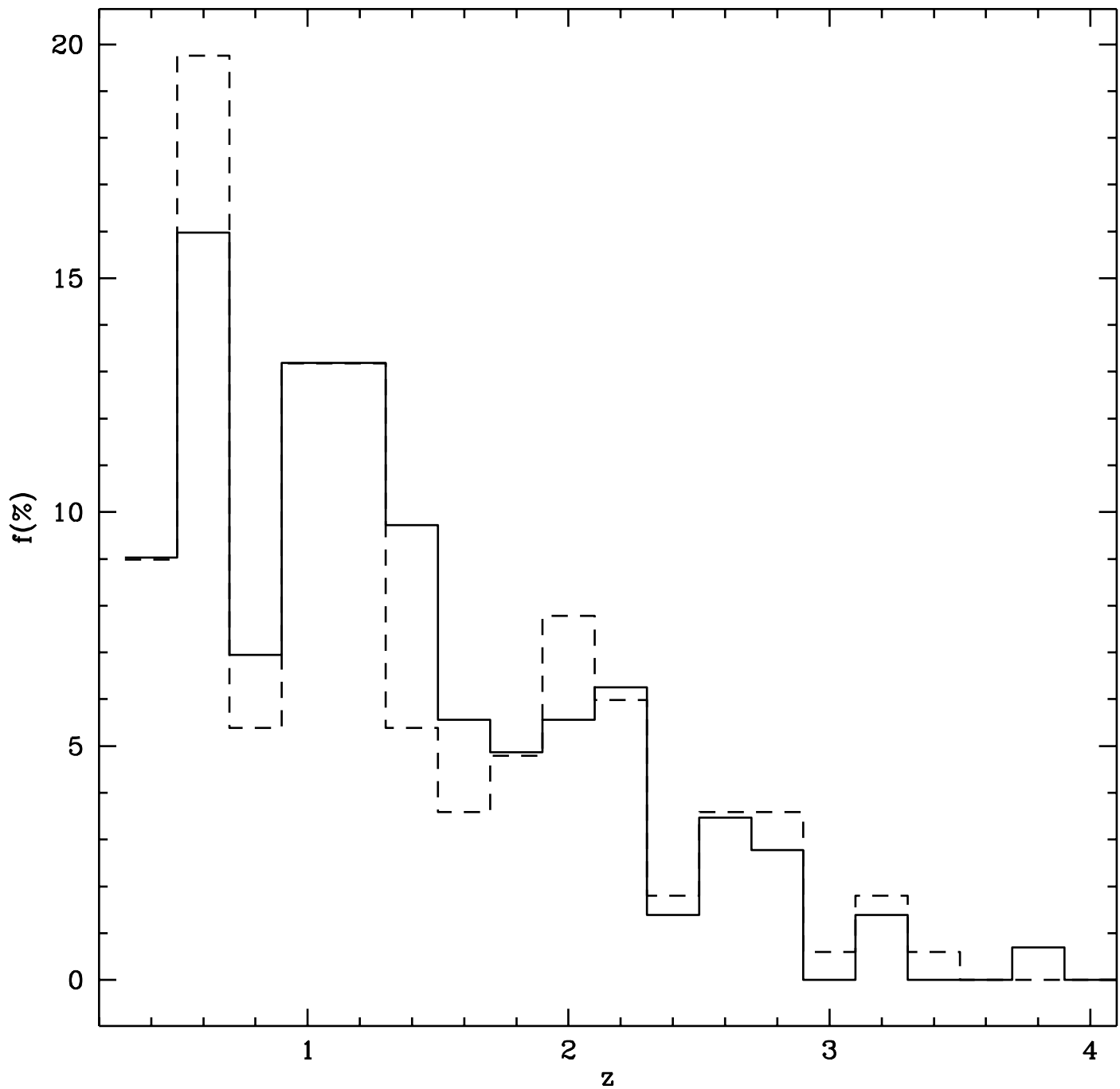
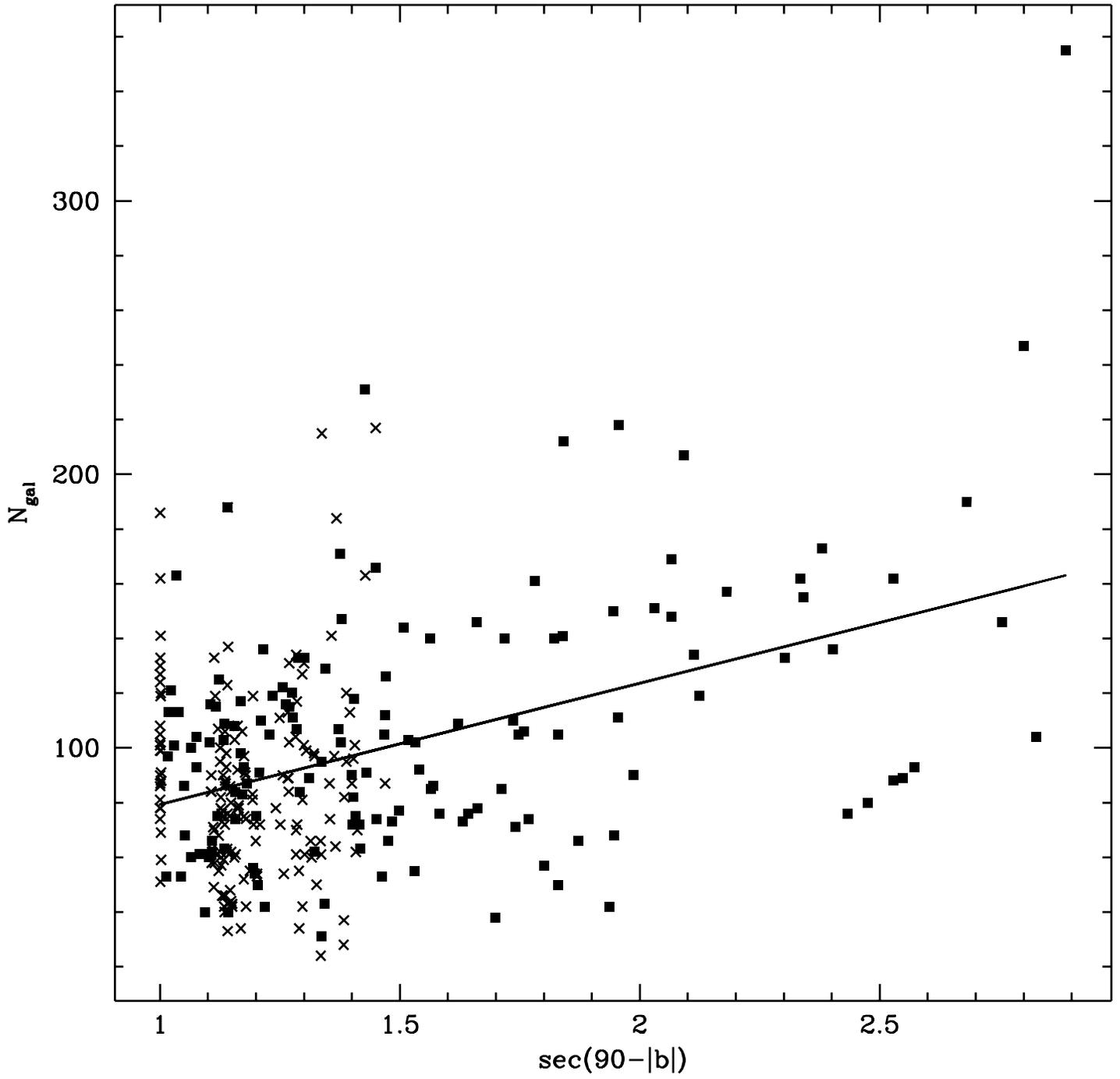
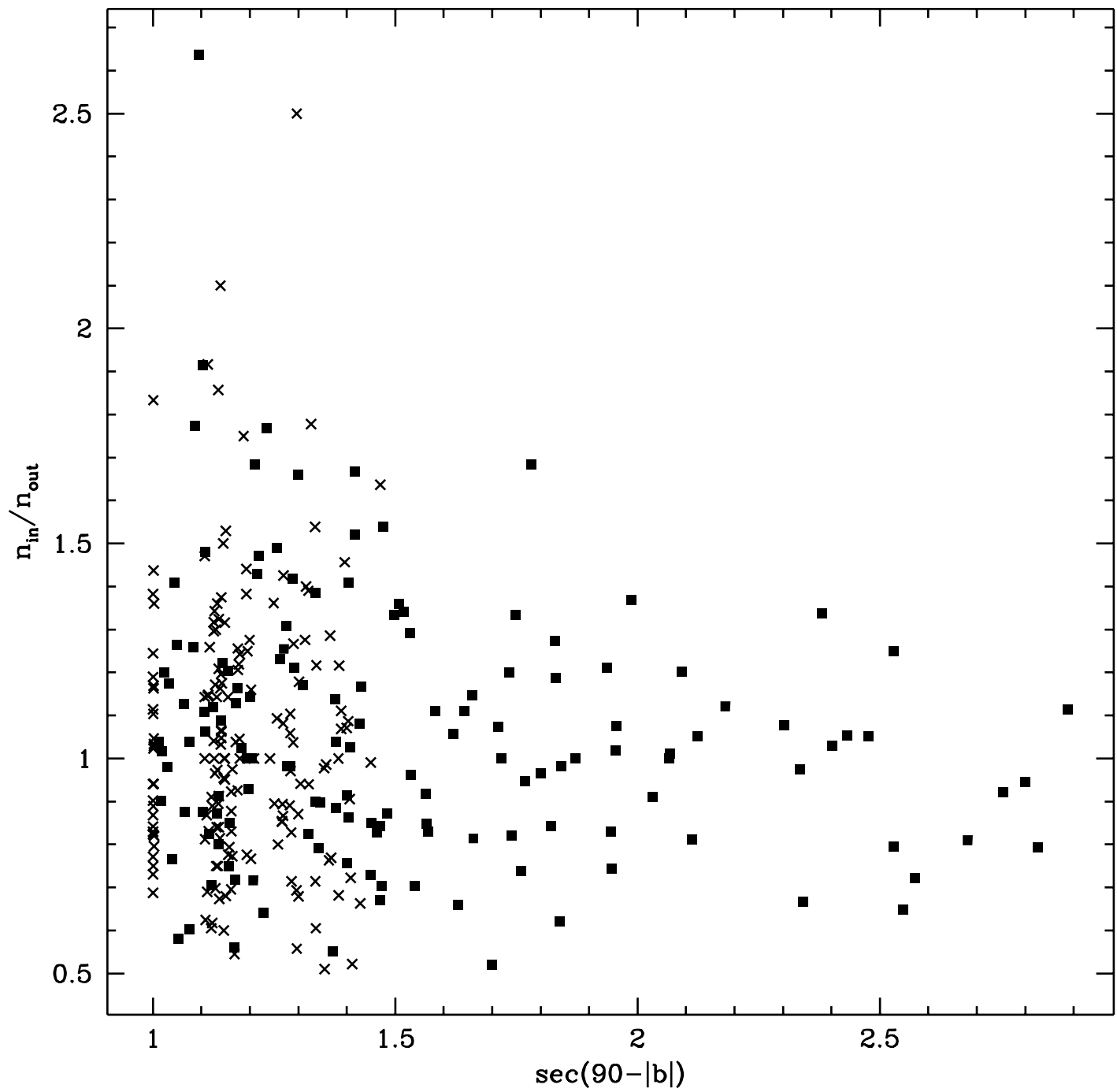
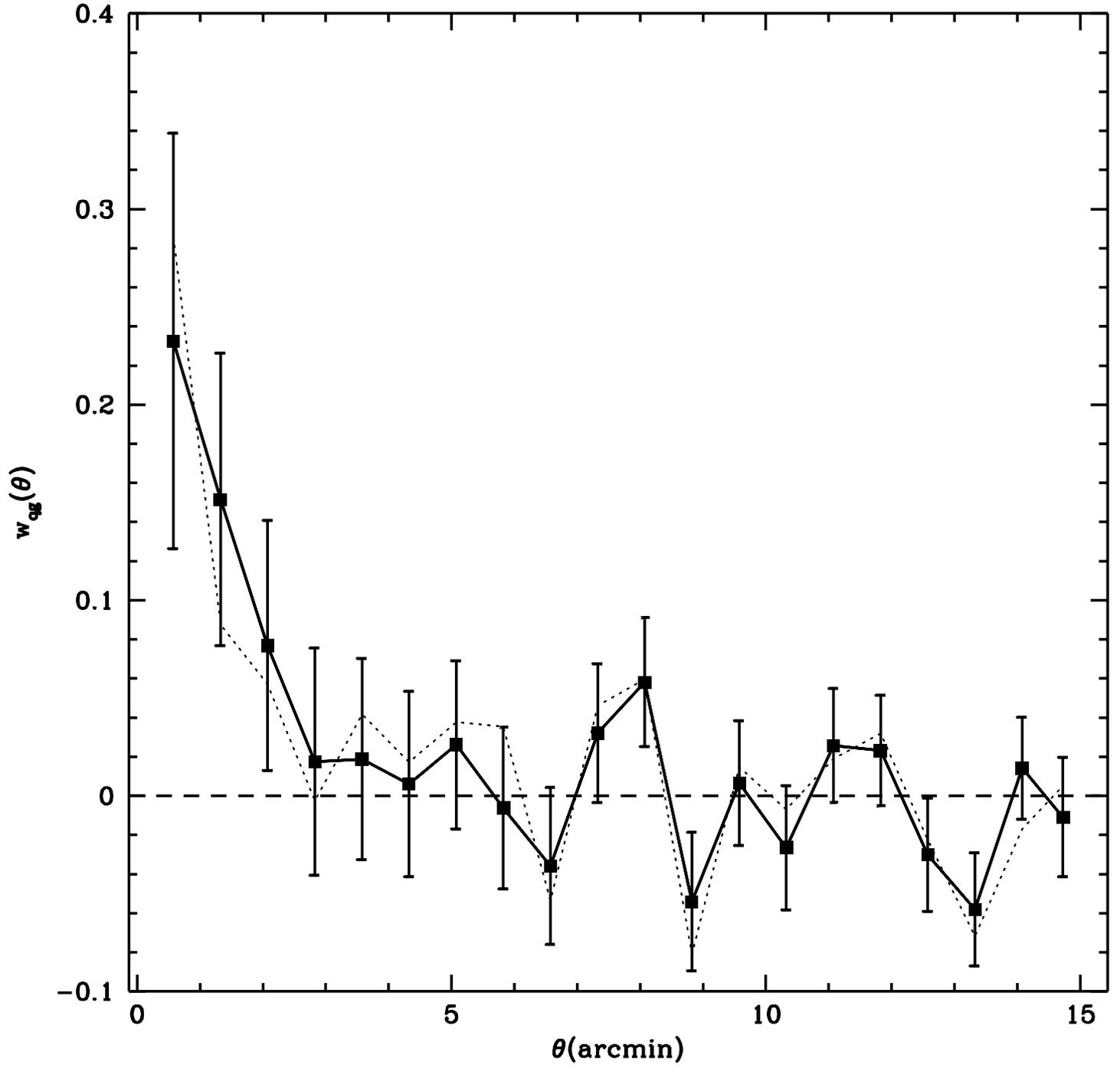
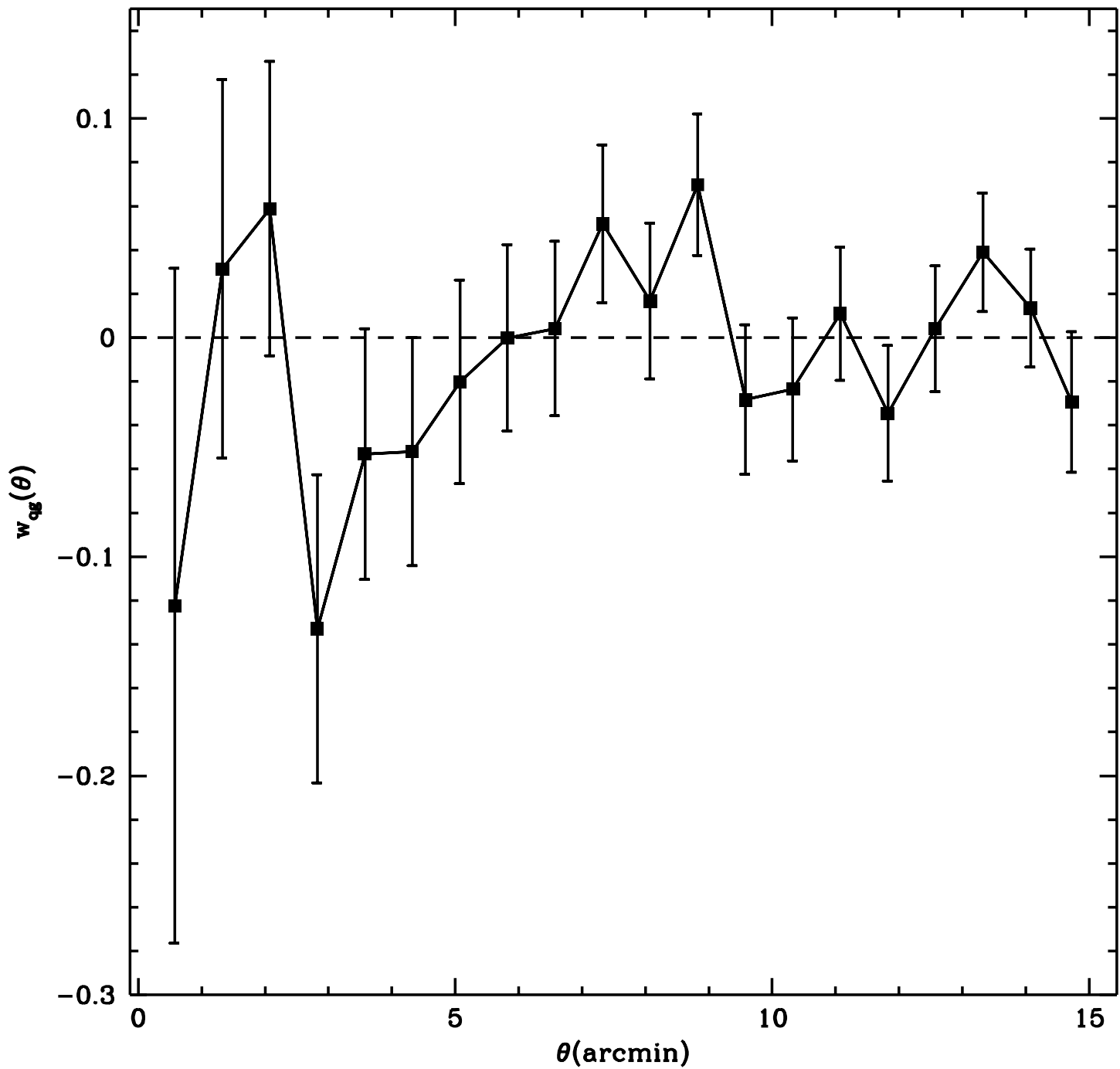
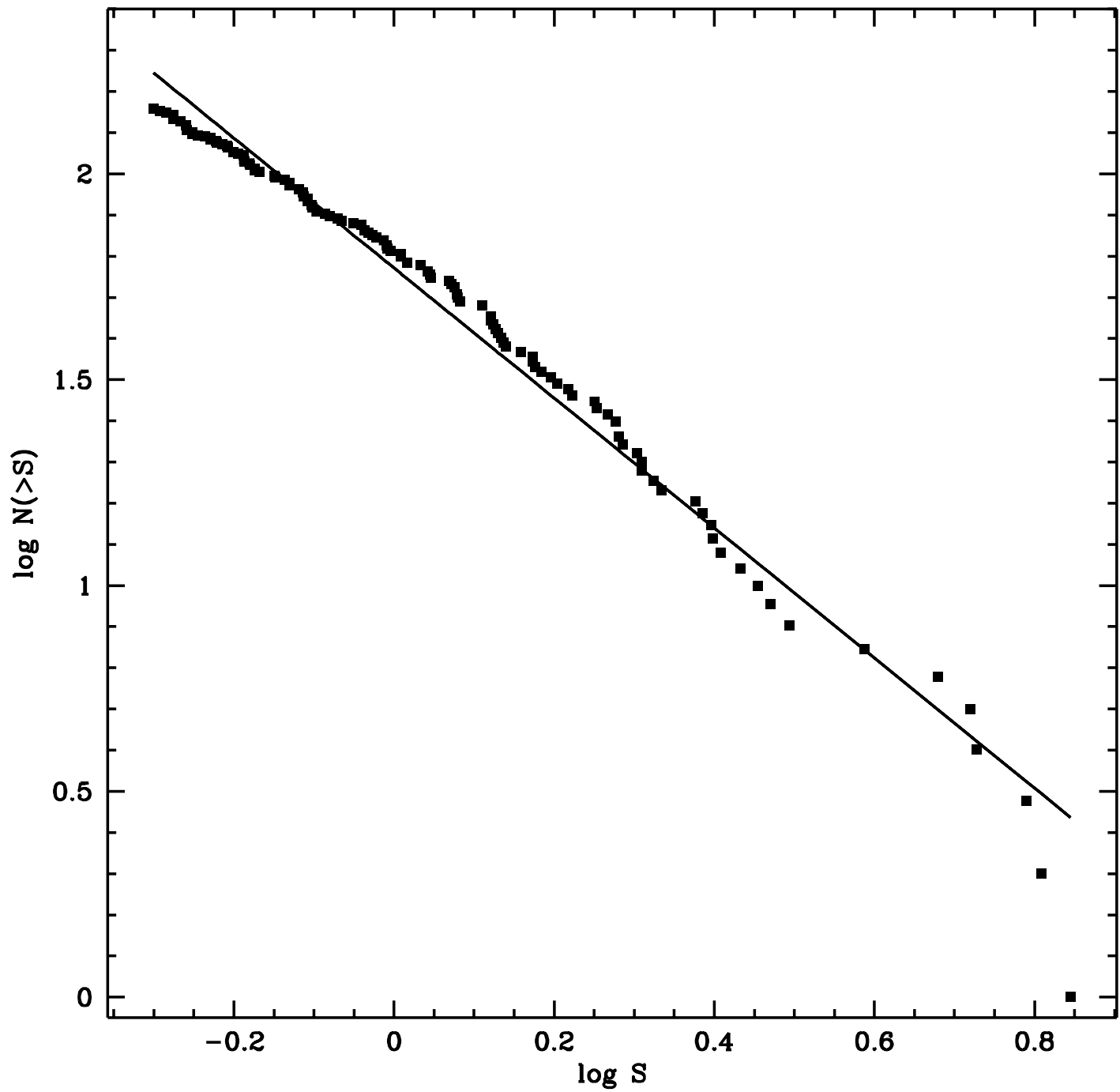
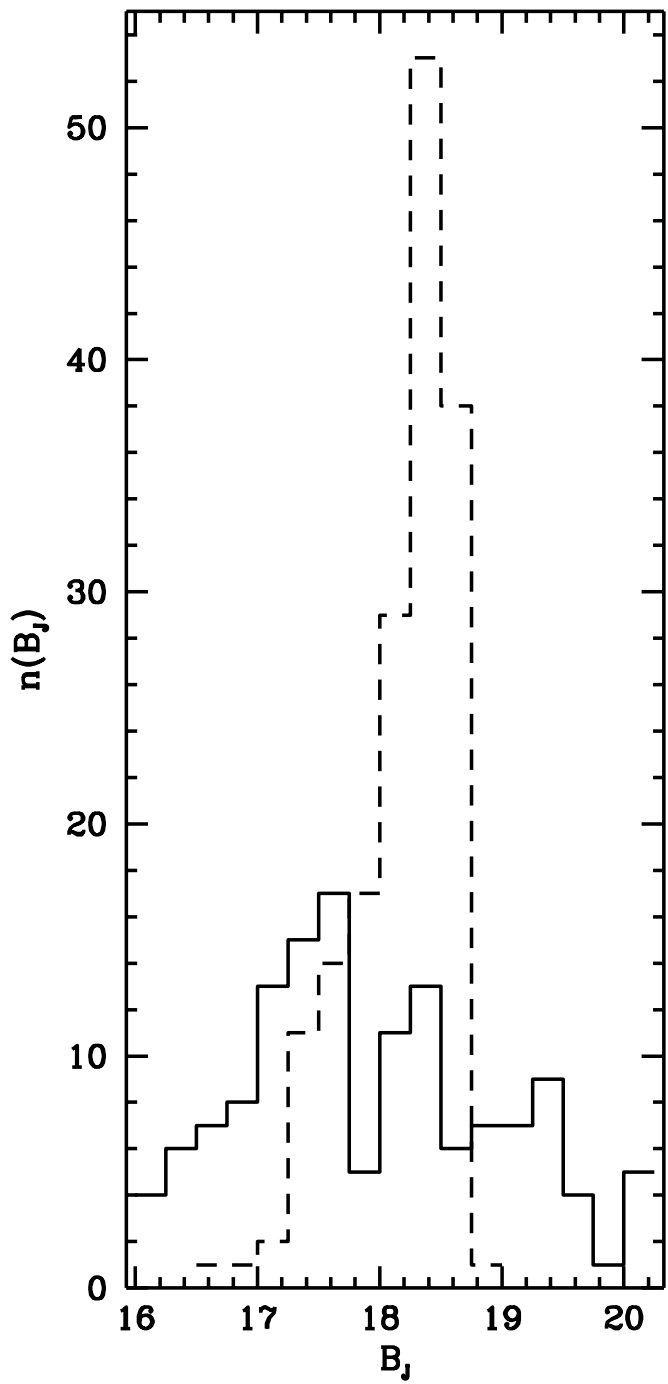
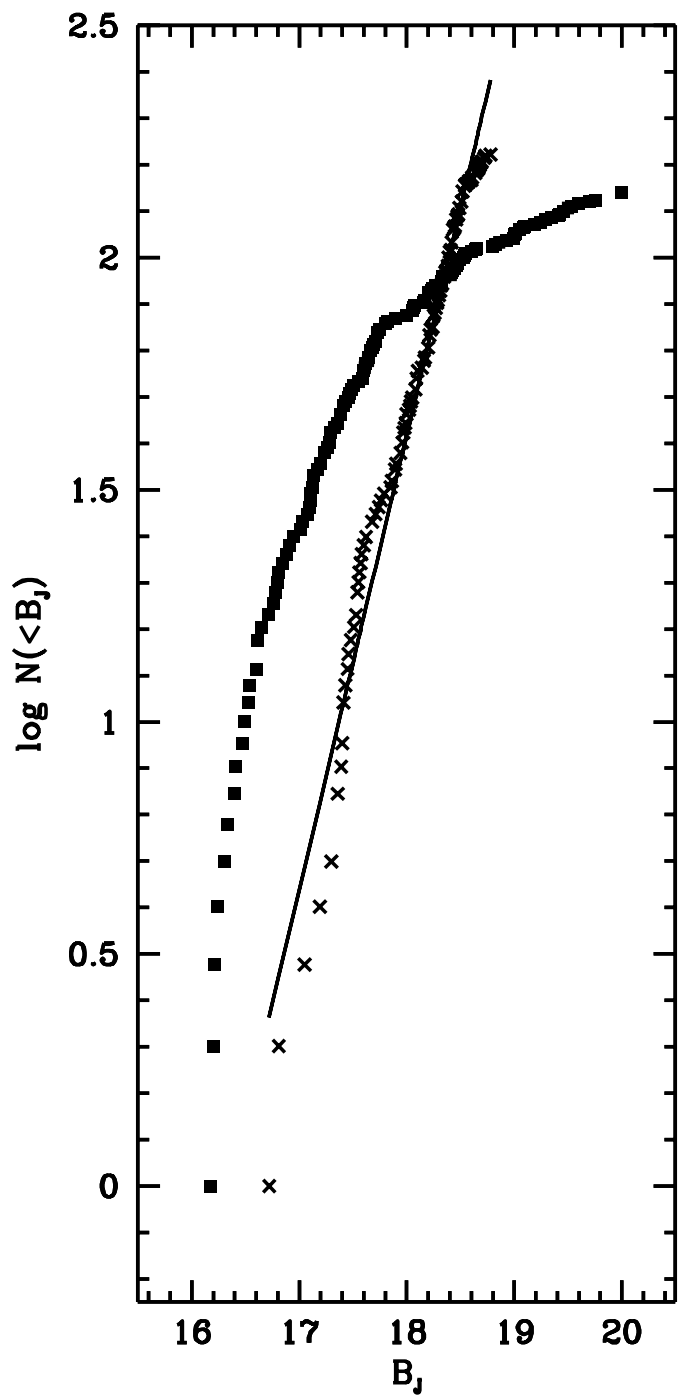
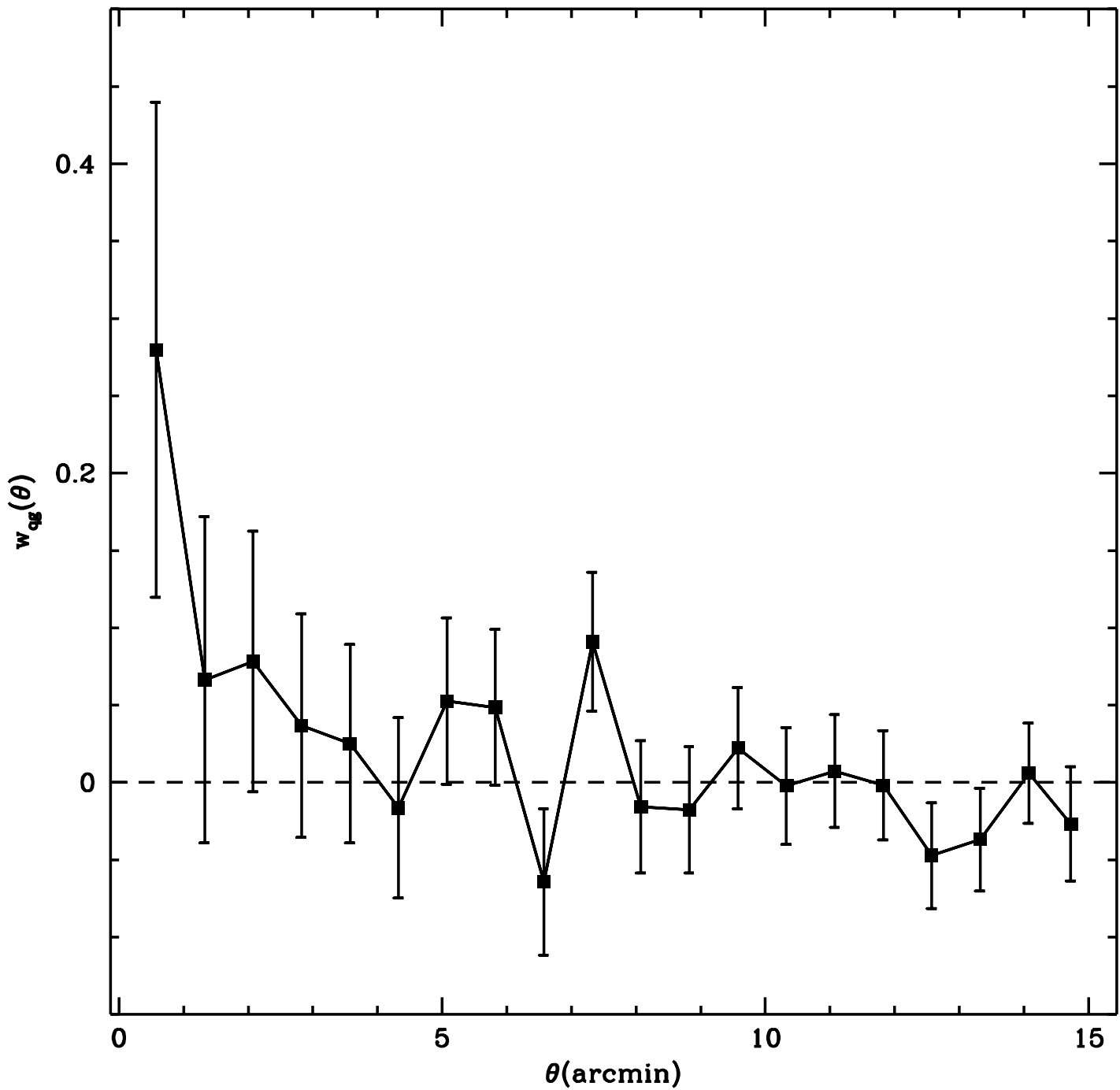
\end{document}